\documentclass[%
 superscriptaddress,
 nofootinbib,
 amsmath,amssymb,
notitlepage
]{revtex4-2}

\usepackage{tablefootnote}
\usepackage[table,x11names,dvipsnames,table]{xcolor}
\usepackage{listings}
\usepackage{siunitx}
\usepackage{graphicx}% Include figure files
\usepackage{dcolumn}% Align table columns on decimal point
\usepackage{bm}% bold math
\usepackage{amsmath,amssymb}
\usepackage[export]{adjustbox}

\begin{document}

%\preprint{APS/123-QED}

\title{Sensitivity to Kaon Decays to ALPs at Fixed Target Experiments}% Force line breaks with \\

\author{Joshua Berger}
 \email{Joshua.Berger@colostate.edu}
\affiliation{%
 Colorado State University, Fort Collins, Colorado 80523
}%
\author{Gray Putnam}
\email{grayputnam@uchicago.edu}
\affiliation{
 University of Chicago, Chicago, IL 60637
}%
\date{\today}% It is always \today, today,

\begin{abstract}
We study the sensitivity of fixed target experiments to hadronically-coupled axion like particles (ALPs) produced in kaon decays, with a particular emphasis on current and upcoming short-baseline neutrino experiments. We demonstrate that below the kaon decay mass threshold ($m_a < m_K - m_\pi$) kaon decay is the dominant production mechanism for ALPs at neutrino experiments, larger by many orders of magnitude than production in pseudo-scalar mixing. Such axions can be probed principally by the di-photon and di-muon final states. In the latter case, even if the axion does not couple to muons at tree level, such a coupling is induced by the renormalization group flow from the UV scale. We reinterpret prior results by CHARM and MicroBooNE through these channels and show that they constrain new areas of heavy axion parameter space. We also show projections of the sensitivity of the SBN and DUNE experiments to axions through these channels, which reach up to a decade higher in the axion decay constant beyond existing constraints. DUNE projects to have a sensitivity competitive with other world-leading upcoming experiments.
\end{abstract}

\maketitle

%\tableofcontents

\section{Introduction}

The axion is a well-motivated solution to the Strong CP problem.  Quantum Chromodynamics (QCD) admits a CP-violating coupling
\begin{equation}
    \mathcal{L} = - \frac{\theta\, g_s^2}{8 \, \pi^2} \, G^a_{\mu\nu} \, \widetilde{G}^{a\,\mu\nu}.
\end{equation}
While this coupling can be rotated away by rephasing the quark fields, the anomaly in this rotation then leads to a phase in the quark mass matrix that is physical provided all the quarks as massive, as current data indicates.  This phase then contributes to the neutron electric dipole moment, leading to a constraint
\begin{equation}
    \overline{\theta} = \theta + \arg \det [Y_u \, Y_d] \lesssim 10^{-10}.
\end{equation}
There is no symmetry reason for this quantity to be small, so the explanation for why it is small remains a theoretical puzzle.

One elegant solution to this puzzle is a Peccei-Quinn QCD axion~\cite{Peccei:1977hh}.  In this model, one introduces a new field, the axion, that couples to $G \, \widetilde{G}$ and renders the parameter $\theta$.  The phase of the axion field is a massless Nambu-Goldstone boson of a spontaneously broken Peccei-Quinn symmetry under rephasing the quarks.  As QCD becomes strongly coupled, the axion develops a potential that dynamically relaxes the effective $\overline{\theta}$ to 0.  It also lifts the mass of the axion away from zero, as the Peccei-Quinn symmetry is anomalous under QCD.  This model has been studied extensively, particularly the low mass version with $m_a \sim 10^{-5}~\text{eV}$ where the axion can also be a viable dark matter candidate.  

This model suffers, however, from an effect called the axion quality problem.  Gravity is not expected to respect any global symmetries, including the Peccei-Quinn symmetry.  It can lead to additional explicit breaking of the symmetry beyond the anomaly, distorting the low energy potential of the axion and shifting $\overline{\theta}$ away from zero.  Unless this shift is less than $10^{-10}$, it would be inconsistent with data and invalidate the solution to the Strong CP problem.

A proposed solution to this quality problem is the heavy QCD axion, in which it is possible to make $f_a$ sufficiently small to have the QCD potential dominate over any gravitational effects and keep $\overline{\theta} < 10^{-10}$.  In order to achieve this, one needs to introduce new ingredients to the axion model, such as an extended gauge group~\cite{Dimopoulos:1979pp,Rubakov:1997vp,Gherghetta:2016fhp,Dimopoulos:2016lvn,Agrawal:2017ksf,Agrawal:2017evu,Gaillard:2018xgk,Lillard:2018fdt,Csaki:2019vte,Gherghetta:2020keg,Gherghetta:2020ofz,Valenti:2022tsc} or a $Z_2$ mirror symmetry~\cite{Berezhiani:2000gh,Hook:2014cda,Hook:2019qoh,Fukuda:2015ana}.

In this work, we consider the sensitivity of fixed target experiments to a heavy QCD axion. It has been shown that high-intensity fixed target experiments are sensitive to heavy QCD axions~\cite{Co:2022bqq,Ema:2023tjg}, and searches have been performed~\cite{CHARM:1985anb,ArgoNeuT:2022mrm,Coloma:2022hlv}.  Much of this work focuses on axion production through mixing of the axion with the light neutral mesons, $\pi^0$, $\eta$, and $\eta^\prime$~\cite{Georgi:1986df,Bauer:2017ris,Aloni:2018vki,Kelly:2020dda,Co:2022bqq}.  This mixing is challenging to calculate consistently~\cite{Bauer:2021wjo,Co:2022bqq}.  
Another channel has been the source of recent study, in which the axion is produced via $K \to \pi \, a$ decays.  For $K^+$ and $K^0_L$ decays, there is an isospin enhancement of this decay, leading to a more significant branching that naively expected~\cite{Bauer:2021wjo,Ema:2023tjg, Cornella:2023kjq}. Prior work has demonstrated this channel can produce a significant kaon decay at rest (KDAR) axion flux at stopped-kaon sources~\cite{Ema:2023tjg}.

As we demonstrate in this paper, kaon decay is the dominant production channel for gluon-coupled heavy axions below the kaon decay mass threshold at neutrino experiments such as MicroBooNE, SBN, and DUNE. This flux orginates from both kaon decays at rest (KDAR) and in flight (KDIF), and searches that combine the two sources will have the greatest reach. As a result, the sensitivity of the future SBN and DUNE experiments to such axions is greater than as indicated by previous studies which focused on the role of axion production through pseudo-scalar mixing~\cite{Kelly:2020dda, Coloma:2023oxx, Co:2022bqq}. We consider the sensitivity of these future experiments, as well as previous searches, to kaon-produced axions through the $a\to \gamma\gamma$ and $a\to\mu\mu$ decay channels. This sensitivity is shown for a ``minimal" axion model where the axion only couples to the Standard Model bosons at the UV scale, as well as a model where the axion couples to muons at tree level. Even if couplings of the axion to gluons are the only couplings introduced at the UV scale, as in a KSVZ-like model~\cite{Kim:1979if,Shifman:1979if}, the axion will develop couplings to photons and leptons by renormalization group evolution~\cite{Chala:2020wvs, Bauer:2020jbp}.  We show that even in this minimal coupling scenario, the di-muon channel has significant sensitivity.

The remainder of this paper is structured as follows.  In section II, we present the axion model that we study and the relevant production and decay modes for our work.  We then survey the experiments that have and will have sensitivity to this model in the region where production in kaon decay is possible in section III.  Our results are presented in section IV.  Finally, we discuss our results and their implications in section V.

% Previous work~\cite{} focused on the possibility of looking for axions produced by KDAR.  We focus on a broader set of kaon decays, with a particular emphasis on cleaner channels such as $a \to \mu^+ \mu^-$.  We demonstrate that this process ($K\to \pi \, a$, $a \to \mu^+ \mu^-$ should have leading sensitivity to the heavy QCD axion. in certain parts of parameter space.

% Additionally, we consider a ``minimal'' version of this model.  Even if couplings of the axion to gluons are the only couplings introduced at the UV scale, as in a DFSZ-like model~\cite{Dine:1981rt,Zhitnitsky:1980tq}, the axion will develop couplings to photons and leptons by renormalization group evolution.  We show that even in this minimal muon coupling scenario, this channel could have significant sensitivity.  Decays of $a\to \gamma\gamma$ also contribute, but are harder to reconstruct experimentally in general and, in most cases, face worse backgrounds.  As such, the muon channel could be an attractive alternative or complement to searches in the diphoton channel, even in this case.

\section{Axion Model}

To model the heavy QCD axion, we do not construct a full UV theory, but rather consider a low-energy effective theory of the form~\cite{Co:2022bqq}
\begin{equation}
    \mathcal{L} = \mathcal{L}_{\text{SM}} + c_3 \, \frac{\alpha_s}{8 \, \pi \, f_a} \, a \, G^{a\mu\nu} \, \widetilde{G}^a_{\mu\nu} + c_2 \, \frac{\alpha_2}{8 \, \pi \, f_a} \, a \, W^{i\mu\nu} \, \widetilde{W}^i_{\mu\nu} + c_1 \, \frac{\alpha_1}{8 \, \pi \, f_a} \, a \, G^{a\mu\nu} \, \widetilde{G}^a_{\mu\nu} + c_\mu \, \frac{\partial_\mu a}{2 \, f_a} \, \overline{\mu} \, \gamma^\mu \, \gamma^5 \, \mu, 
\end{equation}
including a coupling of the axion to muons.  As we note below, one may consider just the gauge boson couplings at the UV completion scale, which will lead to observable $c_\mu$ at the low scale, as well as other subdominant fermion couplings at the mass scale of interest for this work. For this work, we focus on the mass range $m_a < m_K - m_\pi \approx$~\SI{0.36}{GeV}. We discuss the production of decay mechanisms of the axion in this mass range in section \ref{sec:axion-production}, and then detail the specific coupling benchmarks in section \ref{sec:alp-benchmarks}. 

% We focus on the mass window 

% Specifically, 

\subsection{Axion Production and Decay}
\label{sec:axion-production}

\subsubsection{Production from Kaons}

We focus in this work on the production of axions from kaon decay, $K \to \pi a$.  This has been a channel of interest in axion production for a long time~\cite{Georgi:1986df,Bardeen:1986yb} through to today~\cite{Ema:2023tjg}.  The latter work has focused on kaon decay at rest (KDAR) production of axions, but we will highlight that both KDAR and decay in flight (KDIF) are relevant, as is the case for Higgs Portal scalar production in a similar mass range~\cite{Winkler:2018qyg,Batell:2019nwo,NA62:2021zjw,MicroBooNE:2022ctm}. When the axion-gluon coupling ($c_3$) is non-zero, it dominates the kaon decay rate. The branching fraction of kaons into axions can be written as~\cite{Bauer:2021wjo,Bauer:2021mvw}
% Approximate equation neglecting pion and axion mass
% \begin{equation}
%     \frac{\text{Br}(K^\pm \to \pi^\pm \, a)}{\text{Br}(K_S \to \pi^+ \, \pi^-)} \approx \frac{\tau_{K^\pm}}{\tau_{K_S}} \, \frac{f_\pi^2}{8 \, f_a^2}, 
% \end{equation}
% Full equation
\begin{equation}
    \frac{\text{Br}\left(K^\pm \to a + \pi^\pm\right)}{\text{Br}\left(K^0_S \to \pi^+\pi^-\right)} = \frac{\tau_{K^\pm}}{\tau_{K_S}}\frac{2 f_\pi^2 c_3^2}{f_a^2} \left(\frac{m_K^2-m_a^2}{4 m_K^2 - 3 m_a^2 - m_\pi^2}\right)^2 \sqrt{\frac{\lambda(1, m_\pi^2/m_K^2, m_a^2/m_K^2)}{1 - 4 m_\pi^2/m_K^2}}\,,
\end{equation}
where $\tau_{K^\pm}$ and $\tau_{K_s}$ are the charged and short kaon lifetimes respectively and $f_\pi \approx 130~\text{MeV}$ is the pion decay constant.  The $K_S \to \pi^0 \, a$ branching fraction is subdominant, but the for $K_L$, the relation is $\text{Br}(K_L \to \pi^0 \, a) \approx \frac{\tau_{K_L}}{\tau_{K^\pm}}\text{Br}(K^\pm \to \pi^\pm \, a)$~\cite{Bauer:2021wjo, Bauer:2021mvw}.  Notably, these decay channels are isospin favored compared to the dominant kaon decays into pions.  This leads to an enhanced rate of axion production in this channel and, as we will see, leading constraints on axions from this channel.

\subsubsection{Production from Pseudo-scalar Meson Mixing}

Gluon-coupled axions can also be produced through mixing with pseudo-scalar mesons. Although this mechanism is not important for the main result of this paper, we do apply it to benchmark against production in kaon decay. The mixing amplitude is given as the modulus squared of the axion-meson mixing angle $|\theta_{aP}|^2$, where~\cite{Aloni:2018vki, Co:2022bqq}

\begin{equation}
    \begin{split}
        \theta_{a\pi^0} &= \frac{f_\pi}{\sqrt{2}f_a}\frac{1}{6}\frac{m_a^2}{m_a^2 - m_{\pi^0}^2}\\
        \theta_{a\eta} &= \frac{f_\pi}{\sqrt{2}f_a}\frac{1}{\sqrt{6}}\frac{m_a^2 - 4 m_{\pi^0}^2/9}{m_a^2 - m_\eta^2}\,.
    \end{split}
\end{equation}
This expansion breaks down when the axion mass is close to the $\pi_0$ or $\eta$ mass.

\subsubsection{Decays}

In the region of interest, the three pion decay channel of the axion is not open, while decays to lighter fermions such as the electron are generally suppressed even if the relevant couplings are non-zero.  The dominant decays will therefore be to $\mu\mu$ and to $\gamma\gamma$.  The decay rate for these channels is well known, as summarized by Ref.~\cite{Co:2022bqq} for example,
\begin{equation}
    \Gamma(a \to \mu\mu) = \frac{c_\mu^2 \, m_a \, m_\mu^2}{8 \, \pi \, f_a^2} \, \sqrt{1 - \frac{4 \, m_\mu^2}{m_a^2}},
\end{equation}
and
\begin{equation}
    \Gamma(a \to \gamma\gamma) = \frac{\alpha^2 \, |c_\gamma|^2 \, m_a^3}{256 \, \pi^3 \, f_a^2}.
\end{equation}
Here, the coupling $c_\gamma$ needs to be determined at the axion mass scale.  There are several different determinations of this~\cite{Georgi:1986df,Bardeen:1986yb,GrillidiCortona:2015jxo,Bauer:2017ris,Aloni:2018vki,Co:2022bqq} that differ at the $\mathcal{O}(1)$ depending on the chiral perturbation theory approximations being made.  For this work, we apply the result of Ref.~\cite{Co:2022bqq}:
\begin{equation}
    c_\gamma = c_1 + c_2 - c_3\left(1.92 + \frac{1}{3}\frac{m_a^2}{m_\pi^2 - m_a^2} + \frac{8}{9}\frac{m_a^2 -4 m_\pi^2/9}{m_\eta^2 - m_a^2}+ \frac{7}{9}\frac{m_a^2 - 16 m_\pi^2/9}{m_{\eta'}^2 - m_a^2}\right)\,.
\end{equation}
When the axion mass is close to $m_\pi$, $m_\eta$, or $m_{\eta'}$, the mixing formalism breaks down and there is a significant theoretical uncertainty on the coupling.

\subsection{Specific ALP Benchmarks}
\label{sec:alp-benchmarks}

The couplings for the ALP can be fixed in a variety of ways depending on the UV completion of the axion interactions.  A particularly attractive minimal example would be to have vector-like quarks charged under Peccei-Quinn symmetry as well as under QCD, that is the KSVZ model~\cite{Kim:1979if,Shifman:1979if}, or to have the standard model quarks generate the gauge interactions, that is the DFSZ model~\cite{Dine:1981rt,Zhitnitsky:1980tq}.  In addition to these two well-motivated models, we consider the co-dominance model defined by $c_3 = c_2 = c_1$ as a further benchmark, which can easily achieved by appropriate choices of charge for the vector-like quarks that generate the axion couplings to the SM particles.

For the KSVZ model, we would only have $c_3 \neq 0$, while $c_1 = c_2 = 0$ to good approximation at the UV scale where the vector-like quarks are integrated out.  For simplicity, we set our normalization such that $c_3 =1$ at the high scale $\Lambda = 4\, \pi \, f_a$  in all cases.  This choice is arbitrary up to our definition of the scale $f_a$. 
The DFSZ model would require a second Higgs doublet.  It leads to an enhanced muon coupling as we will see.  In this case, at the UV scale, we set $c_3 = 1$ and $c_1 = c_2 = 8/3$.  

As calculated in Ref.\ \cite{Chala:2020wvs, Bauer:2020jbp} and pointed out by \cite{Co:2022bqq,Ema:2023tjg}, couplings to the SM bosons would still result in couplings to all other Standard Model particles at the low scale by renormalization group running to the top quark mass.  For electroweak coupled particles, the running below the top threshold is minimal, as the top quarks dominate the contribution to the running for their couplings to the axion.  We thus run the coupling of the axion to leptons from the high scale to $m_t$ and determine the ``minimal'' coupling of the axion to muons as a benchmark.  This running is somewhat dependent on the details of the UV physics leading to the low-scale interactions we consider, such as the presence of a second Higgs doublet and the dynamics that lead to a viable heavy QCD axion.  We neglect these model-dependent effects in our treatment.  They will have only a small effect on the final result in general.

To determine the effective low-energy couplings, we solve the renormalization group equations outlined in Ref.~\cite{Bauer:2020jbp} assuming each set of $c_1$, $c_2$, $c_3$ at the high scale $\Lambda = 4 \, \pi \, f_a$.  We include the effect of the running SM gauge couplings at one loop and the the running top quark Yukawa coupling, which is by far the dominant correction due to non-zero Yukawas.  We run the Wilson coefficients from the scale $\Lambda$ down to $m_t$.  The corrections to $c_\mu$ below $m_t$ are very small and dominated by heavily suppressed electromagnetic and weak loops, so we take $c_\mu$ at $m_t$ to be the value at the axion mass scale $m_a$.
For the case $c_1 = c_2 = 0$ and $c_3 = 1$, we can validate our results to the numerical result in Ref.\ \cite{Bauer:2020jbp} for $f_a = 1~\text{TeV}$. We find 
\begin{equation}
    |c_\ell(m_t)| = |c_L(m_t) - c_e(m_t)| = 1.8 \times 10^{-4},
\end{equation}
which differs by only about 2\%, likely due to slightly different assumptions about initial conditions. Note that our convention for the gauge boson couplings $c_i$ differs by a factor of 2 compared to their work.

\begin{figure}[!tbh]
\centering
\includegraphics[width=0.49\textwidth, valign=t]{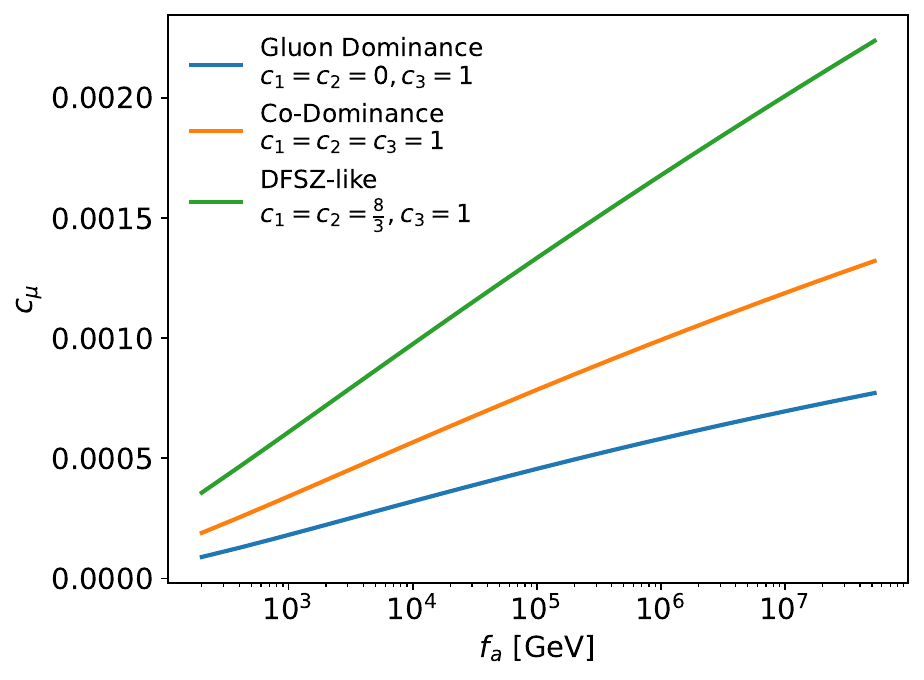}
\includegraphics[width=0.475\textwidth, valign=t]{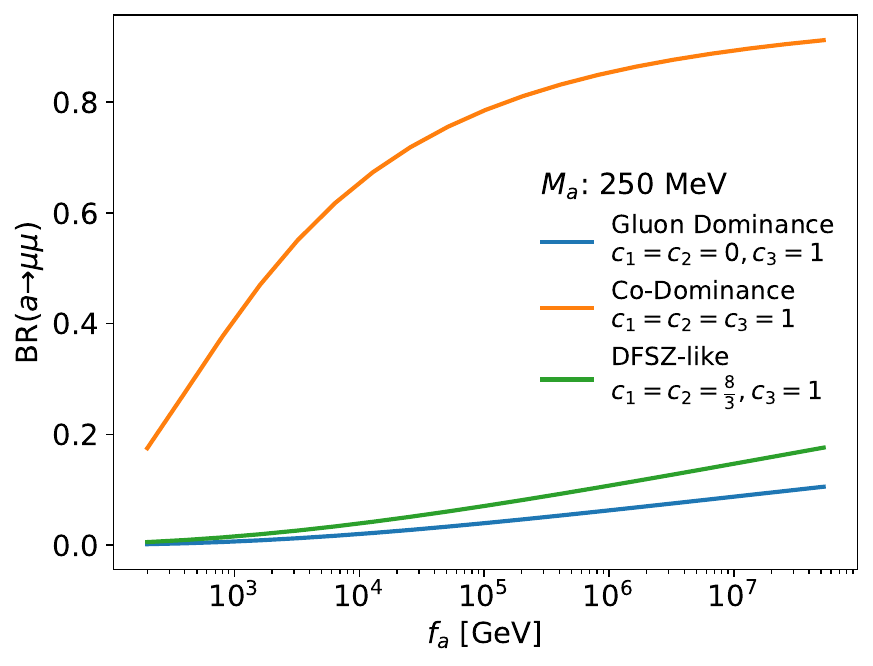}
\caption{(Left) ALP coupling to leptons $c_\ell$ below the electroweak scale assuming $c_\ell = 0$ at the UV scale in the three benchmark ALP models we consider. (Right) Branching ratio of $a \to \mu\mu$ at $M_a = $~\SI{250}{MeV} as a function of $f_a$ for the same three benchmark models.}
\label{fig:minimal-coupling}
\end{figure}

For the model benchmarks we consider, the resulting muon couplings and decay branching ratio at the low scale are shown in Fig.~\ref{fig:minimal-coupling}. The sensitivity of fixed target experiments reaches to $f_a \sim 10^5$~\si{GeV}, an interesting area of parameter space where the decay to muons becomes significant and can even be the dominant decay process; this depends heavily on the choice of the gauge couplings ($c_1, c_2, c_3$), however. Additionally, one may always consider a scenario in which the coupling to leptons at the UV scale is non-zero, opening the window to a far wider range of lepton couplings~\cite{Co:2022bqq}.

\section{Experimental Probes}

\begin{figure}[!tbh]
\centering
  \includegraphics[width=0.49\textwidth, valign=t]{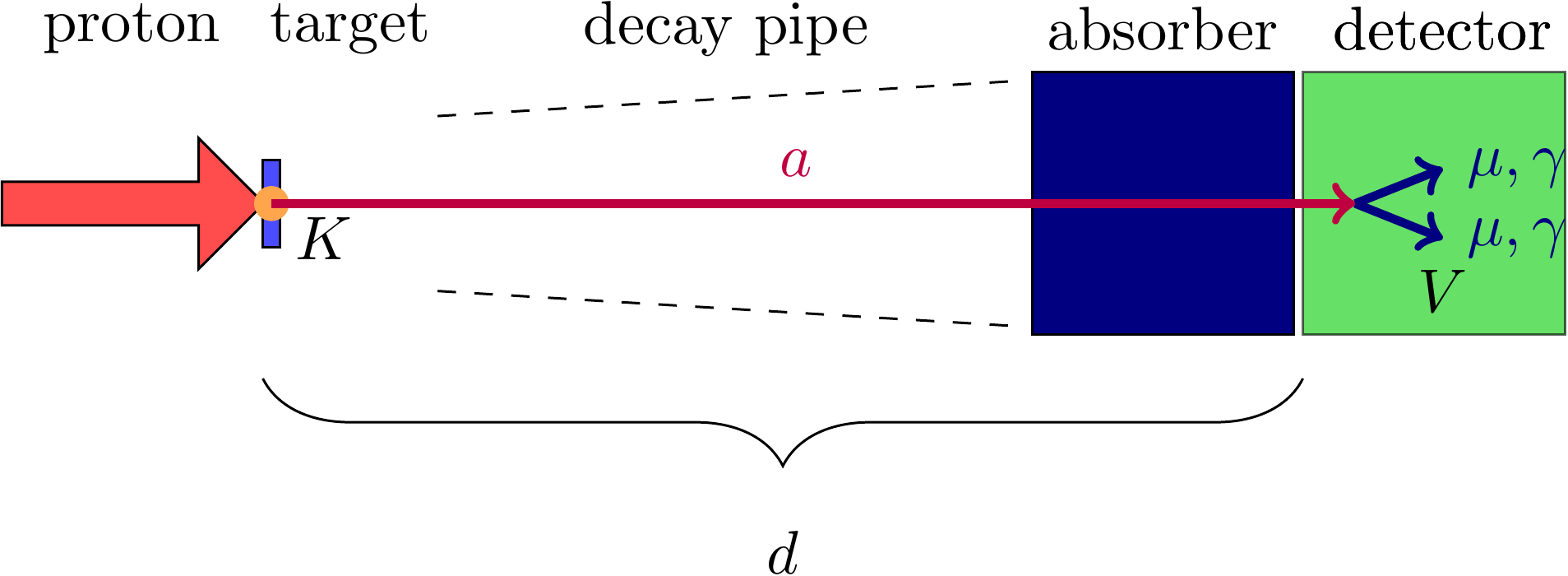}
  \includegraphics[width=0.49\textwidth, valign=t]{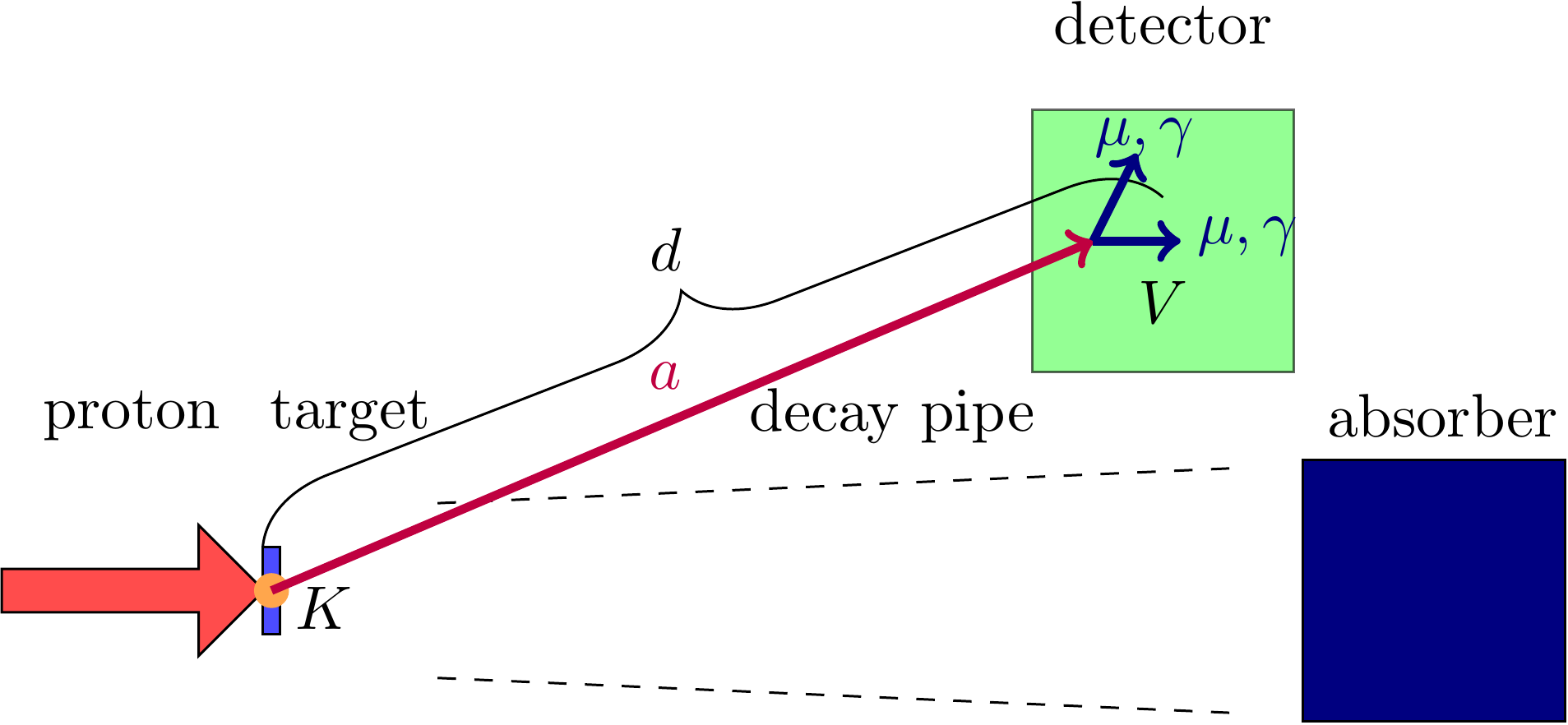}
\caption{Typical setup for an on-axis (left) and off-axis (right) setup at a fixed target neutrino experiment.  A thin target is illustrated, though some experiments have a thick target, leading to most kaons being stopped.  Additionally, for a thin target, some protons produce KDAR in the absorber.}\label{fig:exp-setup}
\label{fig:exp-setup}
\end{figure}

At fixed target experiments, axions are produced in the beam target and travel to a particle detector where they decay into Standard Model particles.  The typical setups we consider for these experiments are illustrated in Fig.\ \ref{fig:exp-setup}. We reinterpret the reach of prior searches at the CHARM and MicroBooNE experiment for the heavy axion model. We also project the sensitivity for the SBN, DUNE, and JSNS$^2$ experiments. With the exception of MicroBooNE, these exclusions require computing the axion rate in each experiment. The detector parameters that enter these computations are detailed in table \ref{tab:detectors}.

\begin{table}
\centering
\newcolumntype{C}[1]{>{\centering\let\newline\\\arraybackslash\hspace{0pt}}m{#1}}
\begin{tabular}{ c|C{3.25cm}|C{4.5cm} | c | c | c| c | c  } 
\hline
 Detector & Active Volume & Fiducialization & Beam & POT & Energy & \begin{tabular}{@{}c@{}}Distance\\ to Beam\end{tabular} & \begin{tabular}{@{}c@{}}Angle\\ to Beam\end{tabular} \\ 
\hline
CHARM & $3 \times 3~\times$~\SI{35}{\meter\cubed} & None. & CHARM & $2.4\times10^{18}$ & \SI{400}{GeV} & \SI{480}{\meter} & \SI{10}{\milli\radian}\\
\hline
 DUNE ND & $7\times 3 \times 5$~\si{\meter\cubed} box + $\pi 2.5^2$~\si{\meter\squared}~$\times$~\SI{5}{\meter} cylinder ($\mu\mu$ only) & \SI{15}{cm} in from sides and front, \SI{1.5}{\meter} in from back of box. Back half of cylinder removed. & LBNF & $1.32 \times 10^{22}$ & \SI{120}{GeV} & \SI{574}{\meter} & 0$^\circ$\\ 
 \hline
 SBND & $4\times 4 \times 5$~\si{\meter\cubed} & \SI{15}{cm} in from sides and front, \SI{1.5}{m} in from back. & BNB & $6.6\times 10^{20}$ & \SI{8}{GeV} & \SI{110}{\meter} & 0$^\circ$  \\ 
 \hline
 ICARUS & $2\times$ $2.97\times 3.17\times 17.9$~\si{\meter\cubed}  & \SI{15}{cm} in from sides and front, \SI{1.5}{m} in from back. & NuMI & $3 \times 10^{21}$& \SI{120}{GeV}  & \SI{800}{\meter} & 5.75$^\circ$ \\
 \hline
 JSNS$^2$ & \SI{20.35}{\meter\cubed} & Included in active volume. & J-PARC SNS & $10^{23}$ & \SI{3}{GeV} & \SI{24}{\meter} & N/A\\
 \hline
\end{tabular}
\caption{Detector configuration parameters for experiments considered in this study.}
\label{tab:detectors}
\end{table}

\subsection{CHARM}

The CHARM experiment used a beam dump of \SI{400}{GeV} protons on a copper target. The detector consisted of a vacuum decay volume instrumented with multiple layers of scintillator hodoscopes and backed by a calorimeter module. The detector sat \SI{480}{\meter} away, 8-\SI{10}{mrad} off-axis from the beam target. CHARM searched for decay vertexes of 2 particles such as $\mu\mu$ and $\gamma\gamma$ as a signature of axions produced in pseudo-scalar mixing~\cite{CHARM:1985anb}. The experiment also published an on-axis search for hadronic final states~\cite{CHARM:1983vkv}. The on-axis search has sub-leading constraints in the mass range we study, and so we do not include it here.
Axions would be produced in CHARM through both kaon decay and pseudo-scalar mixing. The simulation of these two processes is detailed below.

\subsubsection{Kaon Decays at CHARM}

The kaon-induced axion rate in CHARM was estimated with a \lstinline{Geant4}~\cite{GEANT4:2002zbu,Allison:2006ve, Allison:2016lfl} simulation of a \SI{400}{GeV} proton beam impinging an infinite copper target. The large majority of charged kaons (96\%) stop in the target before decaying. $K^-$ are largely captured on copper nuclei, but $K^+$ decay at rest. We find 4.9 kaon decays per proton on target. The probability of a given kaon decay producing an axion decay event (to a final state $X$) in the detector is equal to
\begin{equation}
    p\left(K\to \pi + a(\to X)\right) = \mathrm{BR}(K \to \pi + a) \times \mathrm{BR}(a \to X) \times \int dV \frac{d\Omega'}{d\Omega} \frac{e^{-d/\ell_a}}{4 \pi d^2 \ell_a}\,,
\end{equation}
where the integral ($dV$) is performed over the detector volume, $\frac{d\Omega'}{d\Omega}$ is the Jacobian converting the angular coordinate from the lab frame ($\Omega$) to the kaon rest frame ($\Omega'$), $\ell_a = c\tau_a \beta\gamma$ is the axion decay length in the lab frame, and $d$ is the distance from the kaon decay to the point in the detector volume in the lab frame. This integral was computed with a Monte Carlo simulation program that uniformly sampled a point in the detector volume for each simulated kaon decay.

The CHARM experiment focused on high energy signals characteristic of axion production in pseudo-scalar mixing. Thus, we require the axion to have an energy above \SI{5}{GeV}, a threshold discussed in the CHARM paper. This requirement removes a significant amount ($\sim$~99.5\%), of the axion decays in the decay volume. Above this threshold, di-muon vertices were reported to be identified in CHARM with an efficiency of 85\% and di-photon vertices with an efficiency of 51\%. No events were observed, allowing contours to be set at 2.3 signal events for an exclusion at 90\% CL. The di-muon and di-photon channels were combined into a total exclusion from the more constraining of the two channels at each axion mass.

The CHARM result was previously reinterpreted by Winkler \cite{Winkler:2018qyg} in the context of the Higgs Portal scalar. The phenomenology of the $\mu\mu$ channel is identical for kaon-induced heavy axions and the Higgs Portal scalar. However, our approach differs from that of Winkler in a number of important ways. Winkler computes the kaon rate from primary production on p$-$Cu interactions, whereas we include the amount from re-interactions and therefore find a rate $\sim 5\times$ higher. Winkler includes an attenuation factor from the kaon hadronic interaction length. However, strangeness is conserved in hadronic interactions and so these should not directly attenuate the flux. We compute a morally similar factor from a different perspective: we include kaon energy loss from ionization and hadronic interactions as computed by \lstinline{Geant4}, and put an energy threshold on the resulting axion decay products. We believe that our approach is a robust consideration of the experimental effects relevant to CHARM.

\subsubsection{Pseudo-scalar Mixing at CHARM}

The original search for CHARM interpreted its result through axion-pion mixing~\cite{CHARM:1985anb}. An updated reinterpretation of the search with a modern treatment of axion-pion interactions has been performed~\cite{Jerhot:2022chi}. We re-derive this interpretation to consistently compare the kaon decay and pseudo-scalar mixing production mechanisms for our formalism. Our approach is very similar to the previous reinterpretation, and we find consistent results.

The rate of pseudo-scalar mesons ($\pi^0$ and $\eta$, for the axion mass range we consider) was found from a \lstinline{PYTHIA8}~\cite{Bierlich:2022pfr} simulation of $p-p$ and $p-n$ interactions at the CHARM beam energy (\SI{27.4}{GeV} in the center of momentum frame), scaled by the ratio of protons to neutrons in copper. The simulation was run with \lstinline{SoftQCD::all = on}. Since the pseudo-scalar mesons do not propagate through the target, PYTHIA is adequate to simulate their production. The meson kinematics were turned into axion kinematics by keeping the energy and direction fixed in the lab frame and re-scaling the axion momentum. This choice is arbitrary and violates energy-momentum conservation. We found though that other choices, such as keeping the energy or momentum in the center of mass frame fixed, did not change the kinematics significantly for the energies and masses considered. For each simulated axion, the chord length ($\ell_c$) of the axion ray through the detector volume was computed. Then, the probability of decay for each axion was found according to
\begin{equation}
    p\left(P^0\to (a\to X)\right) = |\theta_{aP}|^2 \times \mathrm{BR}(a \to X) \times e^{-d/\ell_a} \frac{\ell_c}{\ell_a}\,,
    \label{eq:MC-K}
\end{equation}
where $d$ is the mean distance along the chord. As in the case of production from kaon decay, axions were required to have an energy above \SI{5}{GeV}. The sensitivity threshold was set at 2.3 events, for an efficiency of 85\% for di-muon decays and 51\% for di-photon decays. The di-muon and di-photon channels were combined into a total exclusion from the more constraining of the two channels at each axion mass. 

\subsection{MicroBooNE}

The MicroBooNE experiment has published a search for the Higgs Portal scalar produced by kaon decays at rest (KDAR) in the NuMI beam absorber \cite{MicroBooNE:2022ctm}. The heavy axion signal process ($K \to \pi + a(\to\mu\mu))$ is identical to that of the Higgs Portal scalar and therefore this search puts new limits on the axion model. 

Since the flux is from KDAR, it is mono-energetic. Thus, the total rate can be described simply by 
\begin{equation}
    N_\mathrm{sig} = \mathrm{BR}(K^+ \to \pi^+ a) \times \mathrm{BR}(a \to \mu\mu)\times \frac{e^{-d / \ell_a}}{ \ell_a} \times \frac{V}{4\pi d^2}
    \label{eq:MC-P0}
\end{equation}
where $\mathrm{BR}$ is the branching ratio, $\ell_a$ is the axion (or scalar) decay length at the energy of the flux, $d$ is the effective distance to the detector and $V$ is the effective detector volume. We assume that the sensitivity is proportional to the signal rate $N_\mathrm{sig}$ and obtain $V$ and $d$ from the reported exclusion contour in the MicroBooNE result (using the Higgs Portal scalar values of the branching ratio and lifetime). The values depend on the scalar mass. We use the obtained effective volume and distance to compute the equivalent exclusion for the heavy axion model. With this procedure, we are able to compute the reach of the search without making any assumptions beyond the information provided by the measurement.

\subsection{DUNE Near Detector}

\begin{figure}[!tbh]
\centering
\includegraphics[width=0.49\textwidth]{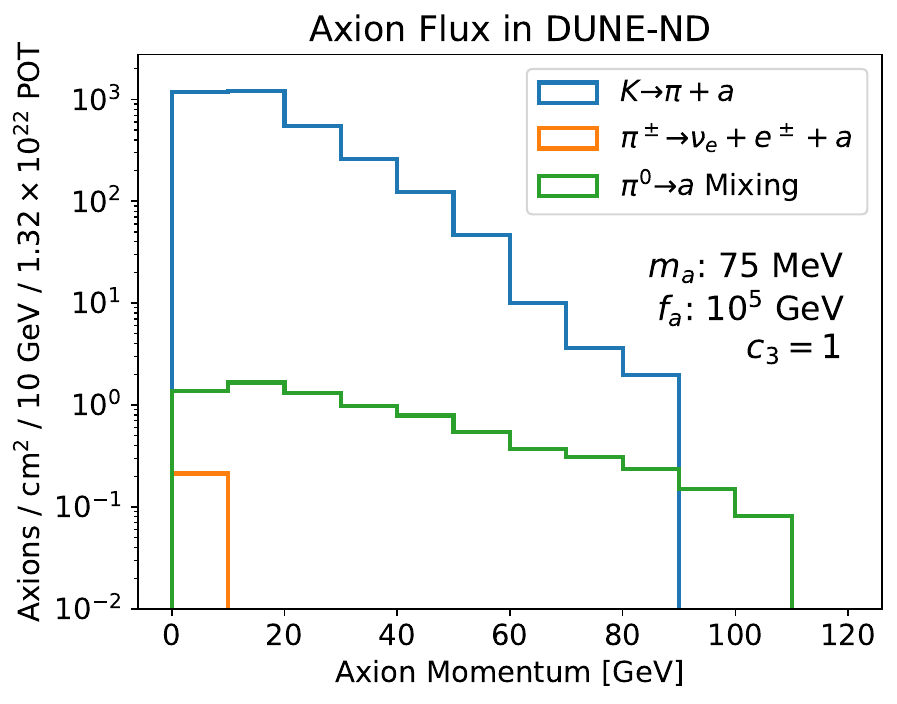}
\includegraphics[width=0.49\textwidth]{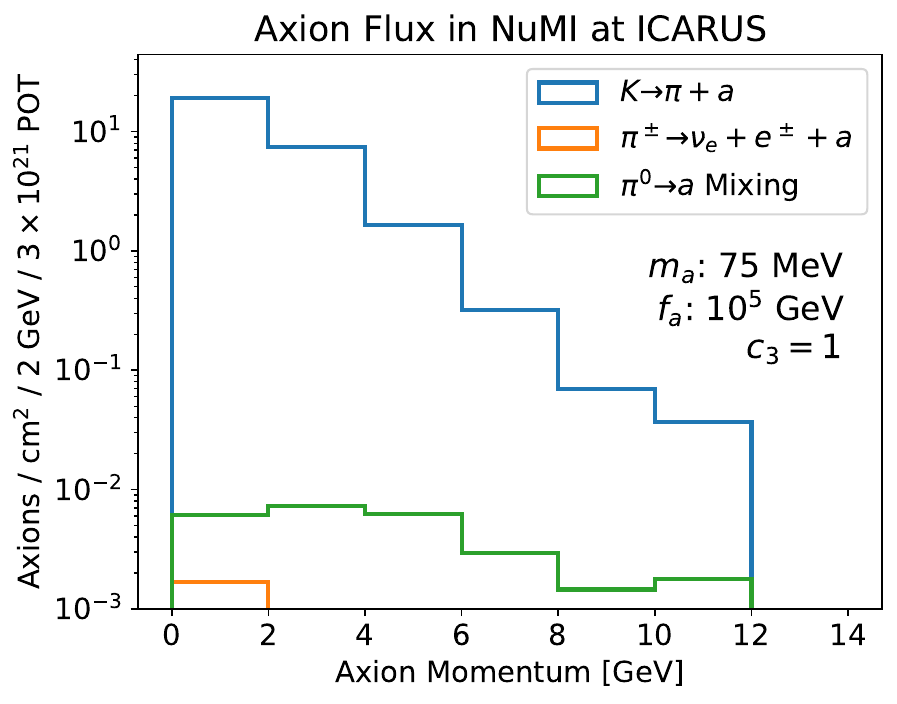}

\caption{Flux of axions at DUNE (left) and ICARUS (right) from production in kaon decay, pion decay, and in mixing with $\pi^0$. The kaon flux includes the contribution from both charged ($K^\pm$) and neutral ($K^0_L$) kaons. The production from kaon decays dominates the flux by many orders of magnitude in both the on-axis (DUNE) and off-axis (NuMI-ICARUS) case.}
\label{fig:DUNE-flux}
\end{figure}

The DUNE near detector complex will consist of multiple neutrino detectors situated \SI{574}{m} downstream of the target hall, where a \SI{120}{GeV} proton beam impinges a graphite target~\cite{DUNE:2020ypp}. As in the case of CHARM, axions can be produced in the target from kaon and pion decay, as well as mixing with pseudo-scalar mesons. The rate from kaon and pion decay was found using the \lstinline{Geant4}~\cite{GEANT4:2002zbu,Allison:2006ve,Allison:2016lfl} based \lstinline{g4lbne} code, which simulates the production and decay of pions and kaons in the LBNF beam. The charged mesons were fed into a Monte Carlo simulation of equation \ref{eq:MC-K} to compute the axion rate. The kaon and pion decay rates were calculated from the weak chiral Lagrangian~\cite{Bauer:2021wjo, Bauer:2021mvw}, as discussed for kaons in section \ref{sec:axion-production}. The pseudo-scalar meson rate was found with a PYTHIA simulation of $p-p$ and $p-n$ interactions at the LBNF beam energy (\SI{15.1}{GeV} in the center of momentum frame), scaled to the relative ratio of protons to neutrons in carbon~\cite{Bierlich:2022pfr}. The pseudo-scalar mesons were fed into a Monte Carlo simulation of equation \ref{eq:MC-P0} to compute the axion rate. 

Fig.~\ref{fig:DUNE-flux} shows the axion flux at the DUNE near detector complex from these different production mechanisms at an example axion mass. Production in kaon decay dominates the flux by many orders of magnitude. We therefore focus on the sensitivity of DUNE and other neutrino experiments to axions from kaon decay. Previous phenomenological studies have addressed the sensitivity of DUNE to axions produced in mixing with pseudo-scalar mesons~\cite{Kelly:2020dda, Coloma:2023oxx, Co:2022bqq}. This study supersedes these previous estimates for axion masses below the kaon production threshold ($m_a < m_K - m_\pi$), but the previous studies still hold for higher masses.

Previous phenomenological studies have discussed the experimental considerations of the $\gamma\gamma$ (e.g.\ Ref.~\cite{Brdar:2020dpr, Kelly:2020dda}) and $\mu\mu$ (e.g.\ Ref.~\cite{Co:2022bqq, Batell:2019nwo}) channels as a new physics signal at liquid argon time projection chamber (LArTPC) neutrino experiments such as DUNE. The di-muon channel has a very small intrinsic background (from neutrino tridents~\cite{Ballett:2018uuc}). However, the channel has a non-intrinsic background from neutrino $\mu\pi^\pm$ events from muon neutrino resonant, coherent, and deep-inelastic-scattering interactions. Charged pions cannot be calorimetrically separated from muons in LArTPCs. However, they can be separated when the pions inelastically scatter, which will happen quite often at the energies relevant for DUNE. Furthermore, pions may be distinguishable with the MeV-scale energy deposits from capture on argon nuclei at the track endpoint~\cite{Castiglioni:2020tsu}. There are also differences in kinematics between di-muon decays and neutrino $\mu\pi^\pm$ interactions: the momenta of the two decay muons forms a mass peak, and their summed vector points back to the target, as opposed to neutrinos where the direction is smeared out~\cite{Batell:2019nwo}. The massive axions also arrive delayed in the detector relative to neutrinos, the timing of which can be reconstructed to $\mathcal{O}$(\si{\nano\second}) precision in LArTPCs~\cite{MicroBooNE:2023ldj, MicroBooNE:2019izn}. Since muons are minimum ionizing, it can also be possible to extend the sensitivity of searches by looking for non-fiducial decays \cite{ArgoNeuT:2022mrm}. 
The di-photon channel has a more challenging intrinsic background from $\pi^0$s produced in neutrino interactions. As for the $\mu\mu$ channel, event kinematics and timing can be used in principle to separate neutrino-induced $\pi^0$s from di-photon decays. 

Searches at LArTPC experiments for di-muon and di-photon signals from KDIF will provide necessary information on how well these different techniques can be applied in practice. For this study, in lieu of a detailed background analysis we draw contours at a fixed event number to project the sensitivity of DUNE in an optimistic scenario where the background rejection is strong in practice. The contour therefore provides a target that the experimental analyses can aim for. The scaling of the event rate goes as ($1/f_A^4$), so even (e.g.) doubling the number will only decrease the sensitivity by $\sim20$\%. We draw contours at 5 fiducial events for the $\mu\mu$ channel and 25 fiducial events for the $\gamma\gamma$ channel.  The difference in the required event rate between the two final states reflects the relative challenge of identifying new physics in both channels. 

We assume a baseline of $1.32 \times 10^{22}$ protons-on-target (POT) taken with an on-axis detector configuration~\cite{DUNE:2020fgq}. For the detector complex, we include a LArTPC with dimensions $7\times 3 \times 5$~\si{\meter\cubed}, as well as a cylindrical muon spectrometer about \SI{5}{\meter} in diameter and \SI{5}{\meter} in height~\cite{DUNE:2020ypp}. The exact form of the muon spectrometer in the DUNE ND is still to be determined. For our analysis, the di-muon channel includes both the LArTPC and the muon spectrometer in the fiducial volume, while the di-gamma channel includes only the LArTPC. We make this choice because the muon spectrometer, regardless of its form, should be able to identify di-muon decays. The case of the di-photon channel is less clear. Furthermore, it may be possible to identify di-muon and di-photon decays with the SAND detector, but we do not include it in our estimate. The LArTPC active volume is fiducialized with an inset of \SI{15}{\centi\meter} in the front and sides and \SI{1.5}{\meter} in the back. Only the front half of the cylindrical detector is counted in the fiducial volume. The detector configuration parameters are specified in table \ref{tab:detectors}. 

\subsection{Short-Baseline Neutrino Program}

The Short-Baseline Neutrino (SBN) Program~\cite{MicroBooNE:2015bmn, Machado:2019oxb} consists of multiple LArTPC detectors at the intersection of the Booster Neutrino Beam (BNB)~\cite{MiniBooNE:2008hfu} and the Neutrinos at the Main Injector (NuMI)~\cite{Adamson:2015dkw} beam at Fermilab. The sensitivity to the BNB is dominated by the SBN near detector (SBND), which sits on-axis \SI{110}{\meter} from the target hall. The ICARUS far detector is the most sensitive to the NuMI beam. It sits $5.75^\circ$ off-axis to the beam, about \SI{800}{m} from the target. We use the \lstinline{g4bnb} and \lstinline{g4numi} codes to simulate the BNB and NuMI beams, respectively. For the BNB, we use a baseline of $6.6\times 10^{20}$ POT~\cite{MicroBooNE:2015bmn}. For the NuMI beam, we use $3\times10^{21}$ POT, which corresponds to about 5 years of runtime for typical rates of POT/year. As is shown in Fig.~\ref{fig:DUNE-flux}, production in kaon decay also dominates the axion flux at the off-axis location of ICARUS in the NuMI beam. The same is true for the on-axis location of SBND in the BNB. Both detector active volumes are fiducialized with an inset of \SI{15}{\centi\meter} in the sides and front and \SI{1.5}{\meter} in the back. The detector configuration parameters are detailed in table \ref{tab:detectors}. The considerations for the $\mu\mu$ and $\gamma\gamma$ channels are the same as for DUNE. 

\subsection{JSNS$^2$}

The JSNS$^2$ experiment will consist of two liquid scintillator neutrino detectors exposed to a flux of $\pi$DAR and KDAR neutrinos. The neutrino beam is generated by stopped mesons from a \SI{3}{GeV} proton beam directed at a mercury target~\cite{Ajimura:2017fld, JSNS2:2021hyk}. The sensitivity of JSNS$^2$ to axions from KDAR has already been demonstrated~\cite{Ema:2023tjg}. We re-derive the sensitivity in this work to benchmark the sensitivity of JSNS$^2$ against other experiments for our formalism.

The sensitivity of JSNS$^2$ is driven by its near detector, which is \SI{24}{m} from the target and has a volume of \SI{20.35}{\meter\cubed}~\cite{JSNS2:2021hyk}. We use the same parameters as Ref.~\cite{Ema:2023tjg} to reproduce their sensitivity to the di-photon channel:  10$^{23}$ POT (about 3 years of runtime) and 0.0054 KDAR per POT, with a contour at 5 events. Backgrounds to the di-photon channel arise from beam-induced neutrons, neutrino interactions, and cosmic-ray gammas. At the visible energies relevant for axions ($>$\SI{227}{MeV}), it has been estimated that the neutrino background dominates and is at 
$\sim$2.5 events or fewer~\cite{Jordan:2018gcd}. The $\mu\mu$ channel has not been studied in depth. However, JSNS$^2$ does plan to measure KDAR $\nu_\mu$ CC interactions~\cite{Ajimura:2017fld}, the signal of which is a single muon at a similar energy range. In addition, the di-muon state would have a pair of Michel decays, which should be a powerful discriminant against backgrounds. We thus project the sensitivity of the di-muon channel with a 5 event contour to demonstrate its reach in the case it is viable.

\subsection{Other Fixed Target Experiments}

Prior work has reinterpreted results from the LSND~\cite{Foroughi-Abari:2020gju} and PS191~\cite{Gorbunov:2021ccu} experiments as putting limits on the Higgs Portal scalar, which are also relevant for the $\mu\mu$ axion channel. However, we elect to not include reinterpretations of these two experiments in this work. The LSND result being reinterpreted is a $\nu_\mu$ charged-current interaction search~\cite{LSND:2003mon}, and the straightforward applicability of that result to $\mu\mu$ events is unclear. In the case of PS191, we find that the experimental result provides too little information to perform a dutiful reinterpretation~\cite{Bernardi:1985ny, Bernardi:1987ek}. In addition, the NuCAL experiment also was sensitive to axions produced in pseudo-scalar mixing~\cite{Blumlein:1991xh}. However, its reach is mostly overlapping with that of the CHARM experiment~\cite{Jerhot:2022chi}, so we do not include it in this study.

\section{Results}

\begin{figure}[!tbh]
\centering
\includegraphics[width=0.49\textwidth]{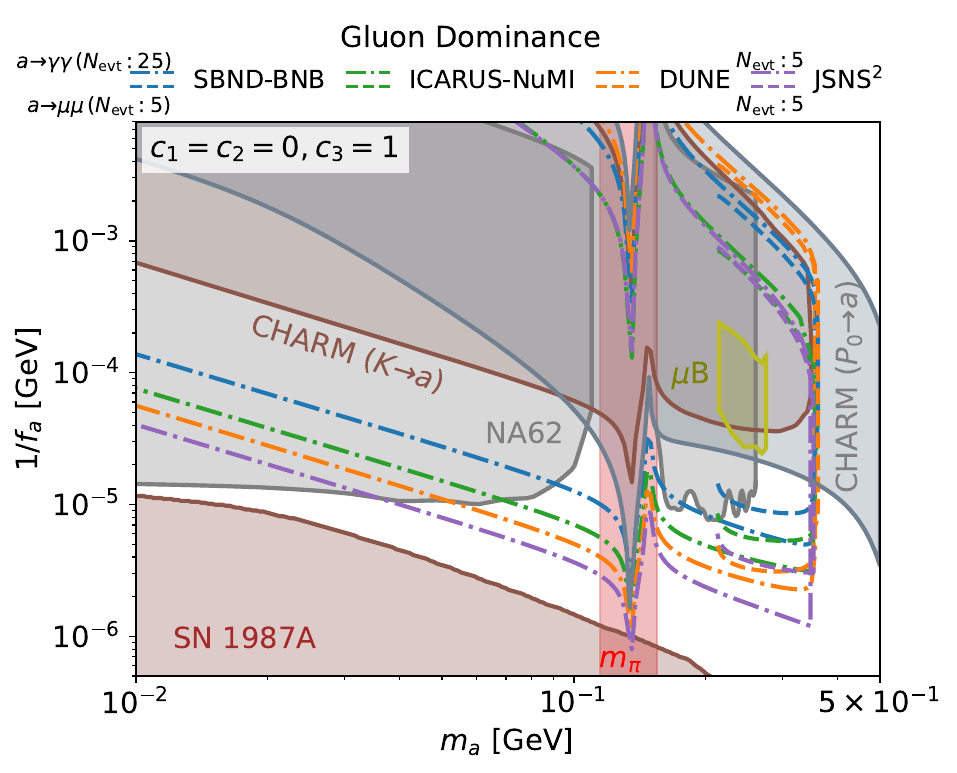}
\includegraphics[width=0.49\textwidth]{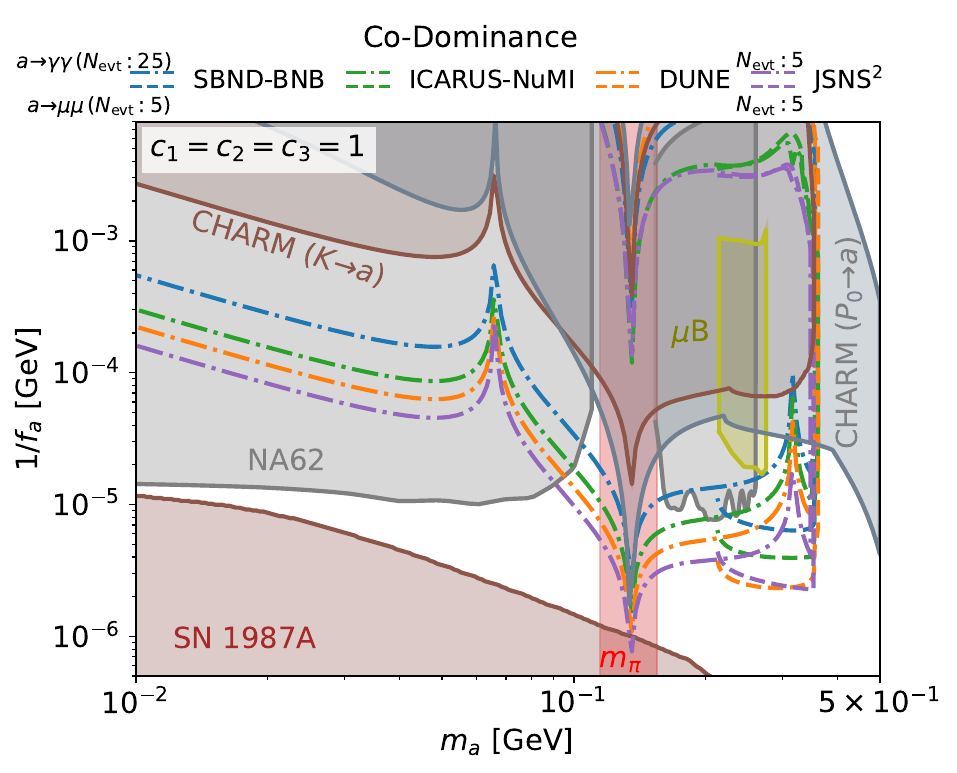}
\includegraphics[width=0.49\textwidth]{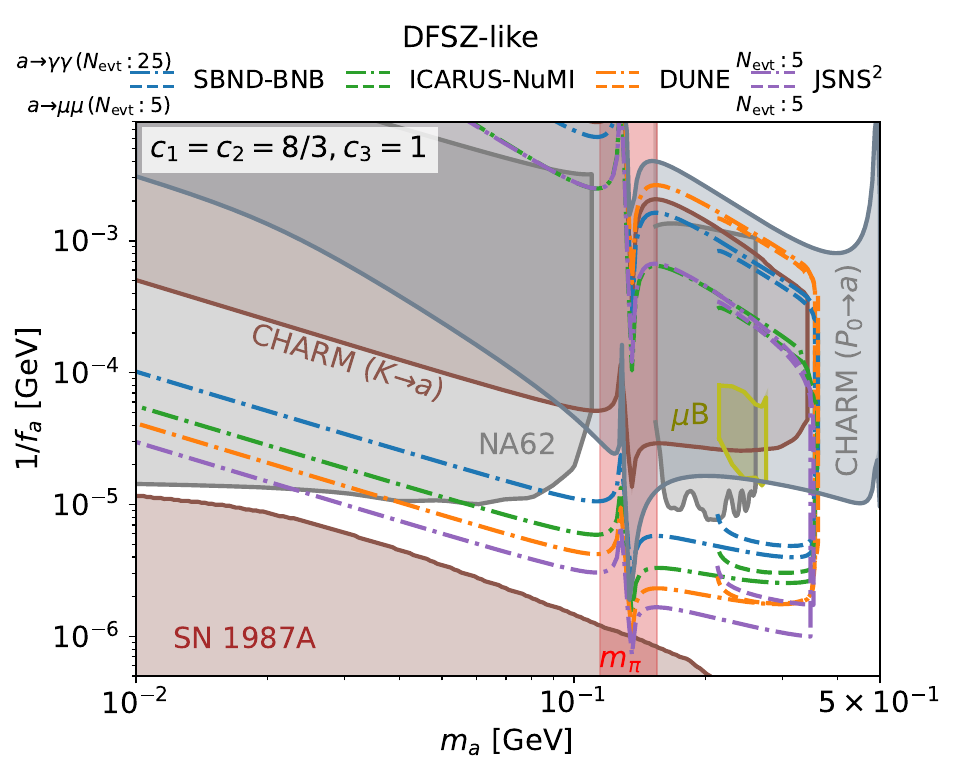}
\includegraphics[width=0.49\textwidth]{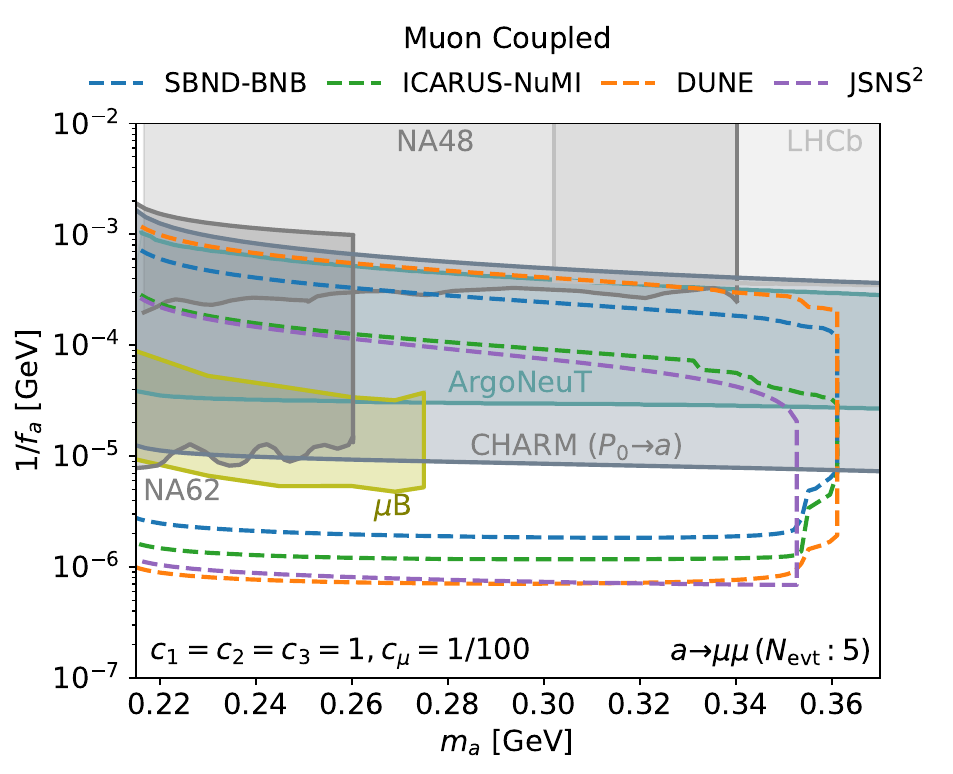}
\caption{Experimental constraints an future sensitivity to the heavy axion model in four coupling scenarios. Excluded regions are shown with filled contours and projections are shown as dashed lines. Our reinterpretation of the CHARM~\cite{CHARM:1985anb} and MicroBooNE~\cite{MicroBooNE:2022ctm} searches are shown alongside a projected sensitivity for future searches at SBN (ICARUS and SBND), DUNE, and JSNS$^2$. The JSNS$^2$ contour is drawn at 5 events for both the di-photon (dot-dashed) and di-muon (dashed) channels. We also include external constraints from NA62~\cite{NA62:2021zjw} and Supernova 1987A~\cite{Ertas:2020xcc}. In the muon-coupled case, constraints from LHCb~\cite{LHCb:2015nkv} and NA48~\cite{NA482:2016sfh} as interpreted by Ref.~\cite{Co:2022bqq}, and ArgoNeuT~\cite{ArgoNeuT:2022mrm} are also relevant.}
\label{fig:result}
\end{figure}

We consider four baseline scenarios for the axion model: gluon dominance, or KSVZ-like ($c_1 = c_2 = 0, c_3=1$), co-dominance ($c_1 = c_2 = c_3 = 1$), DFSZ-like ($c_1 = c_2 = 8/3, c_3=1$), and muon-coupled $(c_1 = c_2 = c_3 = 1, c_{\mu} = 1/100)$. Constraints for each coupling scenario are shown in fig.\ \ref{fig:result}. The choice of the muon coupling strength is such that the axion is long-lived enough to travel the distance to detectors at fixed target experiments, while also still predominantly decaying to muons. This scenario applies the gluon dominance SM boson couplings, but it is not particularly dependent on the values of $c_1$ and $c_2$. 

The projected reach of the SBN and DUNE experiments is computed for both $\mu\mu$ and $\gamma\gamma$ signals for each scenario. The sensitivity from muon decays in the three scenarios only including tree level SM boson couplings originates from the muon coupling generated by the renormalization group flow from the $\Lambda=4\pi f_a$ scale. In the muon-coupled scenario, the projected reach is shown for only the $\mu\mu$ channel, which dominates the sensitivity. The reinterpreted reach of the MicroBooNE $\mu\mu$ search~\cite{MicroBooNE:2022ctm} and CHARM search~\cite{CHARM:1985anb} is also shown. The CHARM result is broken down by production in kaon decay ($K\to a$) and pseudo-scalar meson mixing ($P_0\to a$) so that the reach of two production mechanisms can be compared. Our reinterpretations and projections are compared to other constraints from measurements of kaon decay~\cite{NA62:2021zjw, BNL-E949:2009dza, NA482:2016sfh}, B-meson decay \cite{LHCb:2015nkv}, other fixed target searches~\cite{ ArgoNeuT:2022mrm}, and Supernova 1987A~\cite{Ertas:2020xcc} (which depends heavily on the stellar temperature and density profile \cite{Ertas:2020xcc, Chang:2018rso}). For NA62, we re-compute the sensitivity contours from the information made available in the measurement~\cite{NA62:2021zjw}. We include both the $\gamma\gamma$ and $\mu\mu$ decay channels in the computation of the kaon lifetime. 

The MicroBooNE and CHARM $\mu\mu$ results reinterpreted here both constrain new areas of parameter space in the heavy axion model in the co-dominance and muon coupled scenarios. 
The sensitivity of the CHARM $\gamma\gamma$ channel that we find is consistent with other modern reinterpretations of the search~\cite{Jerhot:2022chi}. Strikingly, the sensitivity of CHARM to axions from kaon production surpasses that from pseudo-scalar mixing at low axion mass. This despite the fact that the CHARM experiment was optimized to search for axions from pseudo-scalar mixing: it was designed with a thick target, and it only searched for high energy signals. The phenomenological enhancement of axion production in kaon decay surpasses even these experimental considerations.

The SBN and DUNE experiments, optimized for kaon decays, all project to constrain significant parameter space in the axion model. It should be emphasized that the event count contours represent estimates of the sensitivity, and that searches will need to be performed in practice to understand just what the reach of the experiments will be. The ICARUS experiment has already been collecting data with the NuMI beam and can perform this search now~\cite{ICARUS:2023gpo}. The sensitivity of DUNE to axions computed here surpasses prior estimations based on pseudo-scalar mixing~\cite{Kelly:2020dda, Coloma:2023oxx, Co:2022bqq} below the kaon decay mass threshold ($m_K - m_\pi$). DUNE's sensitivity is competitive with the previously world-leading projected sensitivity from JSNS$^2$~\cite{Ema:2023tjg}. Searches in the muon channel project to drive the sensitivity above the di-muon mass threshold in the co-dominance and muon coupled scenarios. The sensitivity of the di-photon channel dips in the co-dominance scenario at two zeros in the axion-photon coupling $c_\gamma$: $m_a\sim$\SI{70}{MeV} (caused by $a-\pi^0$ mixing) and $\sim$\SI{320}{MeV} (caused by $a-\eta$ mixing). Treatments of the coupling that neglect mixing with the $\pi^0$ or $\eta$ don't have these zeros and thus obtain different results for the sensitivity of the channel.

\section{Discussion}

The phenomenology of gluonic axions has recently received renewed interest and refinement, especially in the case where its mass is close to the QCD scale. In this work, we have shown that these reevaluations of axion phenomenology open new channels to search for the particle at fixed target experiments: from production in kaon decay (in addition to the traditional pseudo-scalar mixing production), and in decays to di-muon final states (in addition to the traditional di-photon final state). Previous measurements by CHARM and MicroBooNE, when interpreted through these channels, constrain novel areas of parameter space in the heavy axion model. These channels also enhance the projected sensitivity of ongoing and future LArTPC experiments such as SBN and DUNE. In order for these searches to be performed, the strong potential for neutrino background rejection from event kinematics and timing will have to be realized in practice. Investigation at ongoing LArTPC experiments will thus both constrain new parameter space and demonstrate the feasibility of the search for DUNE. 

These promising experimental prospects are matched by recent theoretical work that demonstrate the possibility and desirability of an axion with a mass in the MeV-GeV region. While such a particle would not be a dark matter candidate, it would still solve the strong CP problem while eliding other theoretical issues such as the quality problem. Taken together, the experimental and theoretical prospects demonstrate that the search for axion like particles are a compelling component of the physics program of fixed target experiments with significant potential for discovery. 

\begin{acknowledgments}
We would like to thank Jamie Dyer for helpful discussions and contributions during early stages of this project.  The work of JB is supported by the National Science Foundation under Grant No.\ 2112789. The work of GP is supported by the National Science Foundation under Grant No.\ 2209601, as well as the Grainger Graduate Fellowship at the University of Chicago.
\end{acknowledgments}

\bibliography{ktoalp}% Produces the bibliography via BibTeX.

\end{document}